\documentclass[prl,notitlepage,twocolumn]{revtex4}

\usepackage{amsmath}
\usepackage{amssymb}
\usepackage{bm}
\usepackage{graphicx}
\usepackage{times}

\usepackage{color}
\usepackage{ulem}

\begin{document}

\title{Giant nonlinear response at the nanoscale driven by bound states in the continuum}

\author{Luca Carletti$^1$}
\author{Kirill Koshelev$^{2,3}$}
\author{Costantino De Angelis$^1$}
\author{Yuri Kivshar$^{2,3}$}

\affiliation{$^1$Department of Information Engineering, University of Brescia, INO-CNR Via Branze 38/45, 25123 Brescia, Italy}
\affiliation{$^2$Nonlinear Physics Centre, Australian National University, Canberra ACT 2601, Australia}
\affiliation{$^3$ITMO University, St. Petersburg 197101, Russia}

\begin{abstract}
Being motivated by the recent prediction of high-$Q$ supercavity modes in subwavelength dielectric resonators, we study
the second-harmonic generation from isolated subwavelength AlGaAs nanoantennas pumped by a structured light.
We reveal that nonlinear effects at the nanoscale can be enhanced dramatically provided the resonator parameters are
tuned  to the regime of the bound state in the continuum. We predict a record-high conversion efficiency for nanoscale resonators that exceeds
by two orders of magnitude the conversion efficiency observed at the conditions of magnetic dipole Mie resonance, thus opening the way for
highly-efficient nonlinear metadevices.
\end{abstract}

\maketitle

Meta-optics governed by Mie-resonant nanoparticles has emerged recently as a new field of nanophotonics~\cite{kruk}, and it is expected to complement different functionalities of plasmonic structures in a range of potential applications~\cite{kuznetsov}. All-dielectric nanoresonators have many advantages, including low energy dissipation into heat, that can make them functional building blocks for novel low-loss photonic metadevices. Importantly, the existence of strong electric and magnetic dipole and multipole Mie-type resonances can result in constructive or destructive interferences with unusual beam shaping, and it can also lead to resonant enhancement of magnetic fields in dielectric nanoparticles that bring many novel effects in both linear and nonlinear regimes~\cite{kruk}.

Low-order Mie resonances are known to demonstrate relatively low values of quality factors ($Q$ factors)~\cite{kuznetsov}. Nevertheless, recently it was revealed~\cite{our_prl} that subwavelength nanoscale resonators can support the localized states with high $Q$ factors provided their parameters are closely matched to those of bound states in the continuum (BIC)~\cite{BIC_review} formed via destructive interference of two similar leaky modes~\cite{FW, Sadreev}. A true optical BIC is a mathematical abstraction since its realization demands infinite size of the structure or zero or infinite permittivity~\cite{hsu_2013, alu_2014}. Nevertheless, high-index dielectric nanorods can exhibit high-$Q$ resonances via BIC-inspired mechanism associated with so-called supercavity modes~\cite{rybin}.

\begin{figure}[t]
\includegraphics[width=0.3\textwidth]{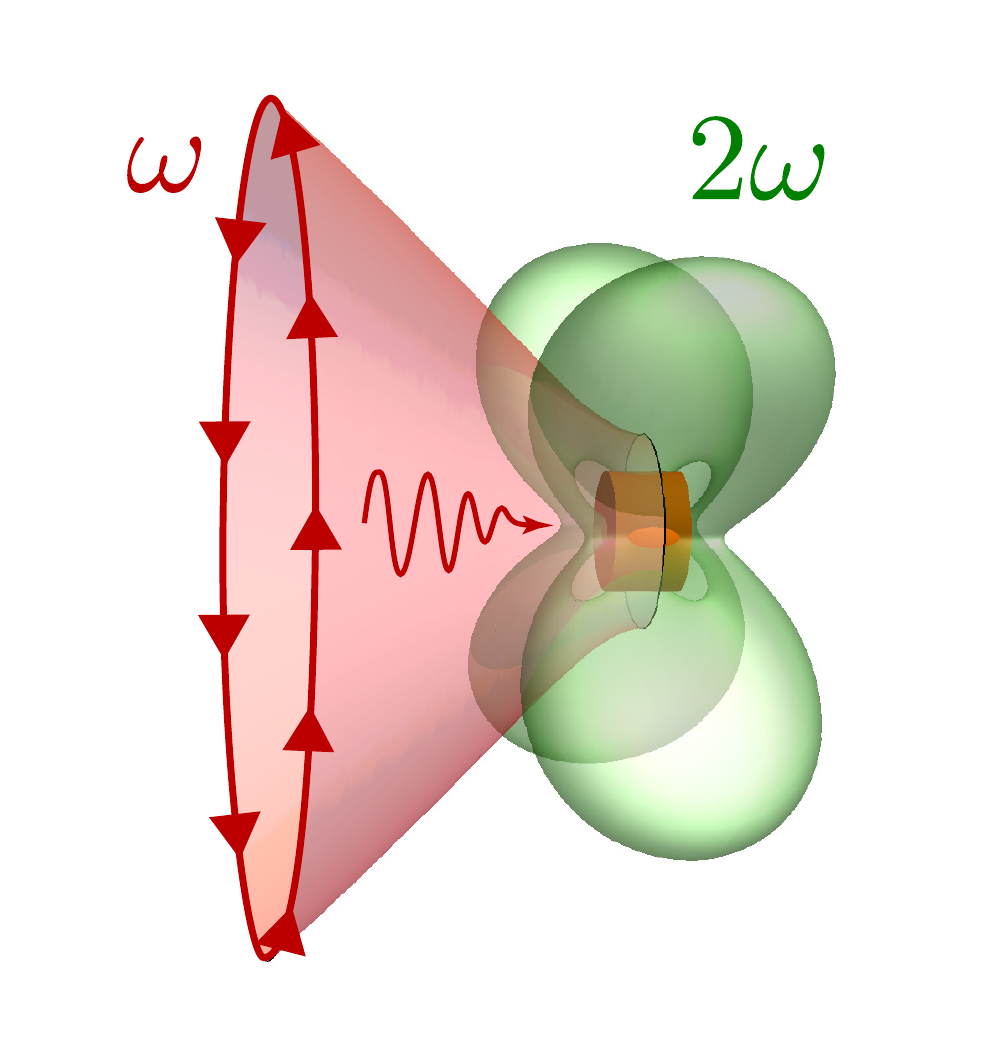}
\caption{Artistic view of the second-harmonic field generated from a subwavelength dielectric nanoparticle pumped by
an azimuthally polarized beam at the conditions of the supercavity mode.}
\label{fig:figure_1}
\end{figure}

The important question now is if those large $Q$ factors can enhance nonlinear effects at the nanoscale~\cite{review,review2} and how large this enhancement can be. In this Letter, we address this question and study the second-harmonic generation (SHG) from isolated subwavelength AlGaAs nanoantennas. We predict a giant enhancement of nonlinear effects provided the resonator parameters are tuned to the regime of the supercavity mode. The predicted record-high conversion efficiency exceeds by two orders of magnitude the largest conversion efficiency at the nanoscale demonstrated so far, opening up the door for many important applications of nonlinear and quantum nanophotonics.

\begin{figure*}[t]
   \centering
   \includegraphics[width=2.0\columnwidth]{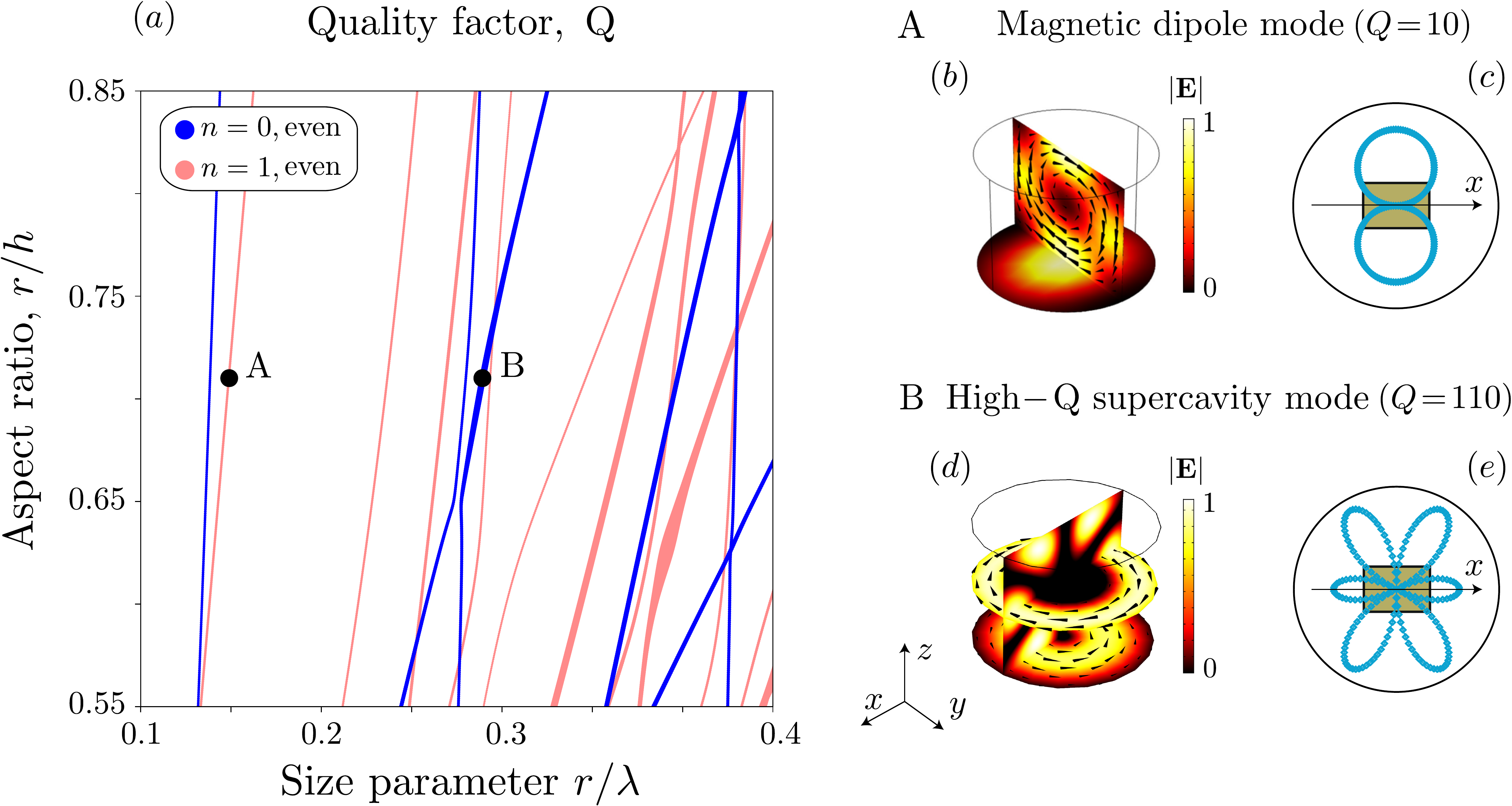} 
   \caption{ Eigenmode spectrum and electric field patterns for ${\rm AlGaAs}$ disk resonator. (a) Eigenfrequencies vs. aspect ratio $r/h$ for even modes with $n = 0$  and $n = 1$ shown with blue and pink dotted lines, respectively. Real part of frequencies is shown by dots. Dot sizes are proportional to the mode Q factors. (b,d) Electric field amplitude and (c,e) directivity pattern of the magnetic dipole mode and high-Q supercavity mode, respectively.}
   \label{fig:figure_2}
\end{figure*}

First, we notice that the nonlinear optics at the nanoscale has many important features, and it is not driven by phase matching~\cite{martti}.
Optical properties of nanoscale structures are governed by strong confinement and resonances~\cite{review}. The field localization
in dielectric nanoparticles is expected to occur near electric and magnetic Mie resonances being driven largely
by their geometric parameters~\cite{kruk}.

Before presenting our main results, we notice that it is customary to define the SHG efficiency as  $\gamma_{\rm SH}=P_{\rm SH}/P_{\rm FF}^2$, where $P_{\rm SH}$ is the total radiated power at second harmonic (SH) and $P_{\rm FF}$ is the pump power incident on the resonator at the fundamental frequency (FF). As defined by this relation, SHG conversion efficiencies in plasmonic nanoantennas up to $5\times 10^{-10}$~W$^{-1}$ have been demonstrated~\cite{But15,Cel15,metall}. An important improvement was recently possible for dielectric nanoantennas where  $\gamma_{\rm SH}\backsim 10^{-6}$~W$^{-1}$ was demonstrated ~\cite{Car15,Lav17,Liu16,Cam16} near multipolar resonances. However, the proposed definition of SHG efficiency does not take into account that only a percentage of the pump power is actually coupled to the cavity mode. To take this issue into account, we define an intrinsic conversion efficiency $\rho_{\rm SH}$ which is independent on the shape of pump beam
\begin{equation}
\rho_{\rm SH} = \frac{\gamma_{\rm SH}}{\eta^2} = \frac{P_{\rm SH}}{\left(  \eta I_m A \right)^2},
\label{eq:rho}
\end{equation}
where $I_m$ is the pump power density impinging on the resonator, and $A$ is the geometrical cross-section of the resonator. The term $\eta$ is the coupling coefficient between the pump beam and the resonator mode at the pump frequency. For resonator modes with Q factors more than $10$, their electromagnetic field is indistinguishable to the profile of modes of the closed cavity since the radiation losses can be treated as a weak perturbation. Therefore, $\eta$ can be defined as the spatial overlap integral between the electric fields of the pump beam, $E_{\rm p}$, and the resonator mode, $E_{\rm c}$, at the upper resonator surface $A$:
\begin{equation}
\eta = \frac{\left|\int_A  E_{\rm p}^*E_{\rm c} dS \right|^2}{\left[\int_A\left|E_{\rm p} \right|^2 dS\right]\cdot \left[\int_A\left|E_c \right|^2 dS\right] }.
\label{eq:eta}
\end{equation}

Thus, the enhancement of extrinsic SHG conversion efficiency $\gamma_{\rm SH}$  can be achieved via increase of efficiency of the coupling between the pump beam and nanoresonator mode and, independently, by enlarging the nanodisk $Q$ factor. On the other hand, when considering structures supporting optical resonances at both SH and pump frequencies, spectral and spatial mode overlap play an important role~\cite{Car15}. While spectral overlap can be improved by tuning pump and SH frequencies to coincide with some of nanodisk eigenfrequencies, spatial overlap involves the mode matching of pump and induced SH polarization fields.


We start our study from the eigenmode analysis. We apply the resonant-state expansion (RSE) method~\cite{4} that allows for careful investigation of eigenmode spectrum of open resonators. We perform RSE calculations for the ${\rm AlGaAs}$ resonator with a fixed permittivity $\varepsilon = 10.73$, which corresponds to material dispersion at pump wavelength $\lambda =1550$ nm (see Supplemental Information for details). We choose the $z$ direction as the disk axis and numerate modes by the radial index, the azimuthal index $n$ and parity  $p=0,1$ with respect to up-down reflection symmetry. Importantly, the RSE analysis for different values of $p$ and $n$ is performed independently and  eigenmodes (resonant states) with $n=0$ are rigorously divided into TE ($\mathbf{E} = E\mathbf{e}_{\varphi} $) and TM ($\mathbf{H} = H \mathbf{e}_{\varphi}$) types~\cite{5}. 

The spectrum of the resonant states (RS) of ${\rm AlGaAs}$ disk resonator with respect to disk aspect ratio $r/h$ for even ($p=0$) modes with $n = 0$ (TE polarization) and $n = 1$ is shown in Fig.~\ref{fig:figure_2}(a) with blue and pink dotted lines, respectively. The RSE method is truncated by $M=16$, $N=452$ for $n=0$ modes and by $M=16$, $N=896$ for $n=1$ modes. Here $N$ is the number of basis modes with frequencies $\omega_{i}$ lying inside a circle of radius $|\omega_{i}R/c|\le M$, where $R$ is the radius of the sphere enclosing the disk.

As can be seen from Fig.~\ref{fig:figure_2}(a), the evolution of disk spectrum with respect to aspect ratio reveals an infinite number of avoided resonance crossings describing strong mode coupling with the formation of high-Q states~\cite{our_prl}, which are the hallmark of non-Hermitian systems with the presence of both internal and external mode coupling~\cite{7}. Here, we focus on two resonant states of the nanodisk, marked as $A$ and $B$ in Fig.~\ref{fig:figure_2}(a). The resonant state $A$ with $p=0, n=1$ represents a conventional magnetic dipole (MD) mode with magnetic moment oriented transversely to the $z$ direction. The resonant state $B$, with TE polarization and $p=0, n=0$, is a high-$Q$ supercavity mode with $Q=110$ associated with the BIC conditions. The electric field and directivity pattern of the MD and BIC modes are shown in Figs.~\ref{fig:figure_2}(b-e), respectively. The electric field of the BIC mode is symmetric with respect to the azimuthal direction which allows to achieve good mode matching with an azimuthally polarized pump and, according to Eq.~(\ref{eq:eta}), to increase the coupling efficiency $\eta$ of the pump into the nanodisk with respect to a linearly polarized pump~\cite{Das15}.

\begin{figure}[t]
   \centering
   \includegraphics[width=1.0\columnwidth]{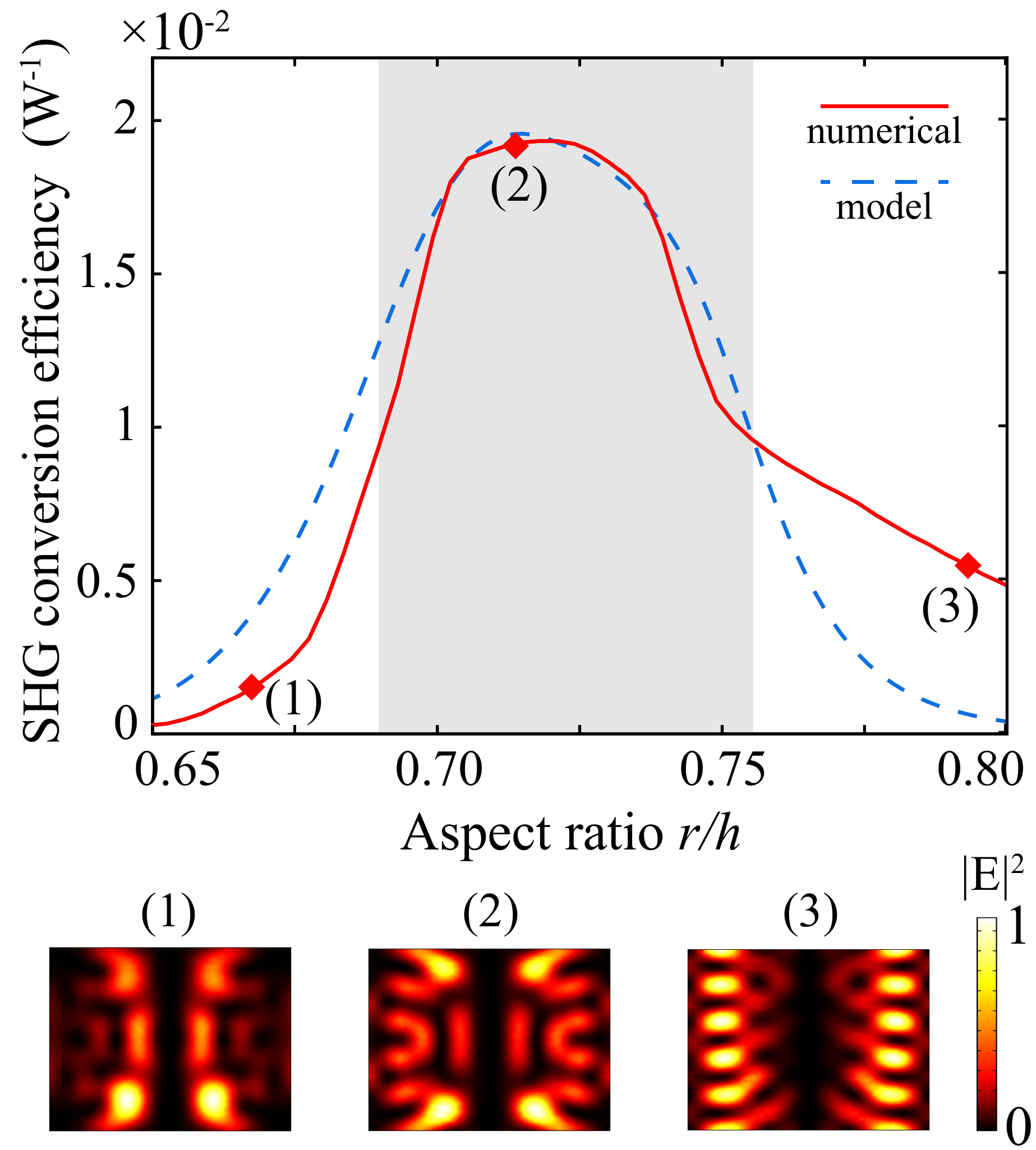} 
   \caption{SHG conversion efficiency $\rho_{\rm SH}$ as a function of the nanoparticle aspect ratio $r/h$ calculated by nonlinear simulations in COMSOL (red solid line) and analytical estimate on the model (\ref{eq:10}) (blue dashed line). Gray area represent the full-width at half-maximum band of the calculated SHG conversion efficiency. Lower inset: Intensity maps at the SH field in a longitudinal disk cross-section for the aspect ratios of 0.675 (1), 0.71 (2), and 0.79 (3).}
   \label{fig:figure_3}
\end{figure}

To investigate the SHG response of the AlGaAs nanoresonator, we use three-dimensional electromagnetic simulations implemented with the finite-element method in COMSOL. The nanodisk is suspended in a homogenous background of refractive index $1$ and the dispersion of the AlGaAs permittivity is fitted from measured values~\cite{Geh00}, with losses taken into account. The nonlinear optical response from the material is estimated assuming an undepleted pump approximation regime. Thus, we use two subsequent simulations: in the first step, the linear optical response of the disk excited at FF is evaluated, while, in the second step, we reproduce the SHG phenomenon by excitation of the nanodisk with the nonlinear currents induced by the FF beam. The induced second-order nonlinear polarizabilities, $P_i^{(2\omega)}$, are deduced from the electric fields in the cavity obtained in the first simulation step (at FF) using the $\chi^{(2)}$ tensor related to a material with the zincblende crystalline structure,
\begin{equation}
P_i^{(2\omega)}=\varepsilon_0 \chi_{ijk}^{(2)}E_j^{(\omega)}E_k^{(\omega)} \quad , \quad i\neq j\neq k
\label{eq:P}
\end{equation}
where $i$, $j$, and $k$ stand for the $x$, $y$, or $z$ axes. For AlGaAs, we use $\chi_{ijk}^{(2)}=100$ pm/V~\cite{Gil16}. Subsequently, we explore two different scenarios. First, we use a focused linearly polarized Gaussian beam with a $60^{\circ}$ angle of incidence with respect to the disk base.  Second, observing the magnetic nature of the BIC resonance,  we use a focused azimuthally polarized beam~\cite{Das15} to selectively enhance optical coupling from the pump beam. The estimated $\rho_{SH}$ for all the analyzed scenarios are listed in Table 1. As can be seen, the BIC-driven SHG conversion efficiency achieves a value as high as $2\times10^{-2}$ W$^{-1}$. As expected, $\eta$ is higher for the azimuthally polarized pump than for the linearly polarized one due to the symmetry of the electric field of the supercavity mode shown in Fig. 2(d). This practically means that for the same incident power on the nanodisk cross-section, using an azimuthally polarized beam results in an increase of about a factor of 100 in the generated SH power $P_{\rm SH}$. Moreover, this aspect is independent on the intrinsic conversion efficiency $\rho_{\rm SH}$, which demonstrates the same order of magnitude for different polarizations of the pump beam.
\begin{table*}
   \caption{Enhancement of the SH response using different combinations of modes at the fundamental frequency and optical excitations. The columns from left to right report the mode description at the FF, the polarization of the electric filed in the pump beam at the FF, intrinsic SH conversion efficiency $\rho_{\rm SH}$ defined in Eq.~\ref{eq:rho}, the coupling coefficient $\eta$, the wavelength corresponding to the FF, and the resonance quality factor Q.}
   \label{tab:tab_1}
\begin{tabular}{|l|c|c|c|c|c|}
  \hline
  \textbf{FF mode}	& \textbf{Polarization}	& \textbf{\boldmath$\rho_{SH}\times 10^4$ (W\boldmath$^{-1}$)}	& \boldmath{$\eta$}	& \textbf{\boldmath$\lambda_{FF}$ (\boldmath$\mu$m)}	& \textbf{Q factor}  \\
  \hline
  BIC		& Azimuthal		& $210$		& $0.77$	& $1.55$	& 110  \\
  \hline
  BIC		& Linear		& $270$		& $0.06$	& $1.55$	& 110  \\
  \hline
  MD		& Linear		& $1.8$		& $0.84$	& $2.98$	& 10  \\
  \hline
\end{tabular}
\end{table*}

As a reference, we compare the enhancement of the BIC-driven SH nonlinear response to that achieved by using the MD resonance at FF, for which many examples in the literature report a very high efficiency (see e.g. Refs. \onlinecite{Mel17,Car17}). Since the observed nonlinear phenomena are volumetric effects, for a fair comparison we kept the disk volume constant. The results from the numeric SHG experiments using the MD mode as FF are summarized in the last row of Table 1. As can be seen, the estimated SH conversion efficiency is of the same order of magnitude compared to previous works, see e.g. Ref. \onlinecite{Car15}, but is smaller by a factor of $100$ than that achieved by using the BIC mode at FF. Such an enhancement can be ascribed to the increase of the $Q$ factor of the mode at FF, that is shown in the last column of Table 1. As expected for second-order nonlinear phenomena, their intensities scale approximately with the factor $(Q/V)^2$, where we use nano definition of the mode volume $V$~\cite{Notomi}. In our case, since the cavity volume is constant while the mode $Q$ factor increases by the factor of $10$ from MD to BIC regime, a rough estimate would indeed suggests that a $100$-fold improvement of the second-order nonlinear interaction should be expected, as found with our numerical simulation results.

\begin{figure}[t]
   \centering
   \includegraphics[width=1\columnwidth]{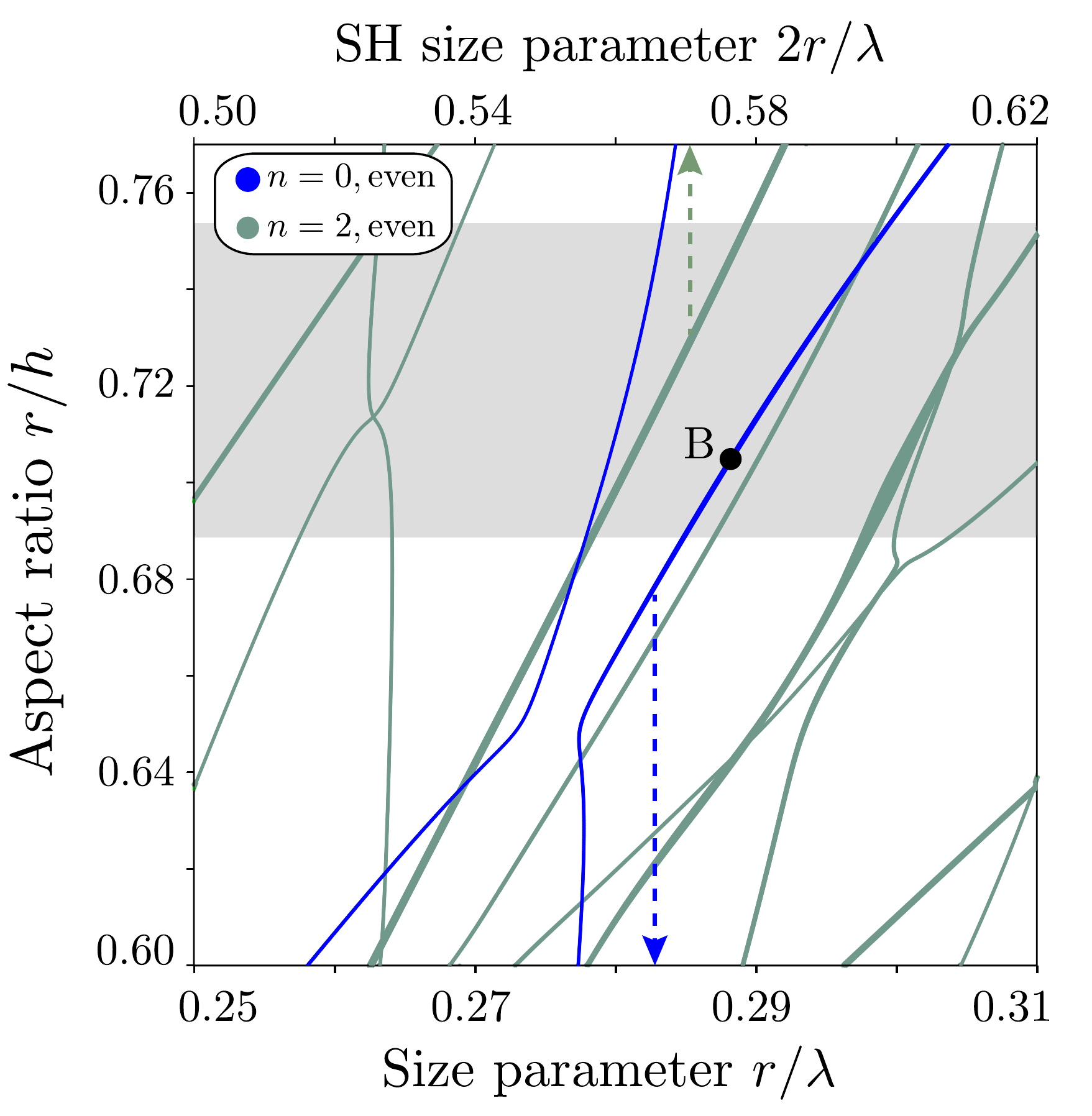}
   \caption{ Eigenfrequencies dependence on disk aspect ratio $r/h$ for the FF modes with the indices $p=0, n = 0$ (TE) and SH modes with $p=0, n = 2$ are shown with blue and cyan dotted lines, respectively. Real part of frequencies is shown by dot positions. The supercavity mode is marked with B as in Fig.2(a). Dot sizes are proportional to the mode Q factor. Pump and SH frequencies are shown in lower and upper horizontal scale, respectively. }
 \label{fig:figure_4}
\end{figure}

To understand the SHG enhancement driven by the BIC-inspired supercavity mode, we analyse the value of $\rho_{\rm SH}$ as a function of the nanodisk aspect ratio by tuning the FF mode on and off the BIC condition. These calculations are performed at a fixed pump wavelength to move exactly along the dispersion line for the mode that is calculated and shown in Fig.~\ref{fig:figure_2}(a). The results depicted in Fig.~\ref{fig:figure_3} (with red solid curve) show that $\rho_{\rm SH}$ reaches its highest value in the region corresponding to the BIC mode condition. The dependence of $\rho_{\rm SH}$ on the nanodisk aspect ratio is, from one side, due to the scaling of the Q factor at FF and, from the other side, due to the resonant response at the SH frequency. Indeed, the insets in Fig.~\ref{fig:figure_3} show that the field profile of the SH response evolves with respect to a change of $r/h$.

To illustrate further this phenomenon, we analyze the coupling between the induced SH polarization and nanodisk eigenmodes applying the RSE at the SH frequency. The electric field of the BIC mode has only one nonzero component $E_{\varphi}^{(\omega)}$, which does not depend on the azimuthal angle exhibiting even symmetry with respect to up-down reflection, thus $E_i(r,\phi,z)=-E_i(r,\phi+\pi,z)$ with $i=x,y$.  From Eq.~(\ref{eq:P}) and the BIC mode symmetry, it follows that only RS of the nanodisk with the indices $p=0$ and $n=2$ are excited at the SH frequency.

Figure~\ref{fig:figure_4} shows the evolution of the nanodisk eigenmode spectrum with respect to the aspect ratio for modes with $p=0, n=0$ (TE) and $p=0, n=2$  calculated by means of the RSE approach.  As in Fig.~\ref{fig:figure_2}, we perform all calculations for the pump wavelength of 1550 nm. The range of aspect ratios for which the calculated SHG conversion efficiency is higher than the half-maximum of calculated the value (see Fig.~\ref{fig:figure_3}) is shown by a gray shadow in Fig.~\ref{fig:figure_4}. Perfect spectral matching between modes takes place when two classes of curves overlap. Figure~\ref{fig:figure_4} shows that there is only one RS at SH frequencies that can be excited by FF in the vicinity of the supercavity mode.

Thus, we apply the temporal coupled mode theory (see Supplemental Information) to estimate the SHG efficiency as a function of the pump frequency $\omega$
\begin{equation}
\tilde\rho_{\rm SH}=C\frac{\gamma_{\rm SH}}{(2\omega-\omega_{\rm SH})^2+\gamma_{\rm SH}^2}\left[\frac{\gamma_{\rm FF}}{(\omega-\omega_{\rm FF})^2+\gamma_{\rm FF}^2}\right]^2,
\label{eq:10}
\end{equation}
where $C$ is a normalization constant, $\omega_{\rm FF}$, $\gamma_{\rm FF}$,  $\omega_{\rm SH}$, and $\gamma_{\rm SH}$ are the FF and SH resonant frequencies and decay rates, respectively. The calculation of $\tilde\rho_{\rm SH}$ for resonant conditions $\omega=\omega_{\rm FF}$ is superimposed to the simulation results in Fig.~\ref{fig:figure_3}, where we observe a good agreement between the two results.

In conclusion, we have presented a new strategy to increase substantially the nonlinear response at the nanoscale by employing the concept of bound states in the continuum realized through supercavity modes. We have predicted that the SHG conversion efficiency in AlGaAs nanoantennas can be increased by at least two orders of magnitude compared to earlier results, due to the mode structure engineering that allows exploiting a high-$Q$ supercavity mode at the fundamental frequency. These results show the great potential of Mie-resonant semiconductor nanostructures for nonlinear nanophotonics, and they constitute a significant step towards the development of highly-efficient frequency conversion metadevices in the subwavelength circuitry.

\begin{acknowledgments}
The authors acknowledge a financial support by the Australian Research Council, the Russian Science Foundation (17-12-01581),
and also the participation in the Erasmus Mundus NANOPHI project, contract number 2013 5659/002-001. They also thank Albert Polman, Anatoly Zayats, Andrey Bogdanov, Bo Zhen, Dragomir Neshev, Harry Atwater, Marin Soljaci\'c, Mikhail Rybin, Sergey Kruk and Yuen-Ron Shen for useful discussions and suggestions.
\end{acknowledgments}


\newpage
\appendix

\section{SUPPLEMENTAL INFORMATION}

In the Supplemental Information we (i) present the rigorous algorithm of implementation of the resonant state expansion for dielectric cylindrical resonators and (ii) derive the equations to estimate the efficiency of second harmonic generation using the temporal coupled-mode theory.

\subsection{Resonant state expansion for dielectric cylindrical resonators}

We calculate the spectrum of complex eigenfrequencies of a dielectric cylindrical resonator by applying the resonant state expansion~\cite{4}, which represents the rigorous method of characterization of spectrum for open electromagnetic systems. We enclose the cylindrical resonator by a homogeneous dielectric sphere with the same value of permittivity. As the first step, we find the eigenmodes (resonant states) $\mathbf{E}_\alpha^{(0)}$ and eigenvalues $\omega_\alpha$ of the sphere solving the Maxwell's equations with the boundary conditions of outgoing waves
\begin{equation}
\nabla\!\times\!\nabla\!\times\!\mathbf{E}_\alpha^{(0)}(\mathbf{r})=\varepsilon(\mathbf{r})\frac{\omega_\alpha^2}{c^2}\mathbf{E}_\alpha^{(0)}(\mathbf{r}).
\end{equation}
Here $\varepsilon(\mathbf{r})$ defines the dielectric sphere enclosing the cylinder and $\alpha=i,n,l$, where $i$ is the radial index, $n$ is the azimuthal index and $l$ is the orbital index. Importantly, the spectrum of eigenmodes is discrete. Next, we use $\mathbf{E}_\alpha^{(0)}$ as the basis and expand the resonant states (RS) $\mathbf{E}_{j}$ of the cylindrical resonator over them
\begin{equation}
\mathbf{E}_j(\mathbf{r})=\sum\limits_{\alpha}{b_\alpha^j}\mathbf{E}_\alpha^{(0)}(\mathbf{r}),
\end{equation}
Here mode number $j=k,n,p$, where $k$ is the radial index, $n$ is the azimuthal index $p=0,1$ is the index of parity of mode with respect to up-down inversion symmetry of the cylinder $\mathbf{E}_j(x,y,-z)=(-1)^p\mathbf{E}_j(x,y,z)$. Remarkably, cylinder RS are not characterized by any fixed orbital index $l$.

Resonants states $\mathbf{E}_j$ can be found as a solution of the non-Hermitian eigenvalue problem~\cite{4}
\begin{equation}
\nabla\!\times\!\nabla\!\times\!\mathbf{E}_j(\mathbf{r})=\left[\varepsilon(\mathbf{r})+\delta\varepsilon(\mathbf{r})\right]\frac{\Omega_j^2}{c^2}\mathbf{E}_j(\mathbf{r}),
\end{equation}
where $\delta\varepsilon(\mathbf{r})$ is a perturbation of permittivity that transforms a sphere into an inscribed cylinder. The problem is reduced to the matrix equation
\begin{align}
&\frac{1}{\omega_\alpha}\sum_\beta\left[\delta_{\alpha\beta}+V_{\alpha\beta}\right]b^j_\beta = \frac{1}{\Omega_j}b^j_\alpha,\\
&V_{\alpha\beta} = \frac{1}{2}\int d\mathbf{r}\ \delta\varepsilon(\mathbf{r})\mathbf{E}_\alpha \mathbf{E}_\beta.
\end{align}
Here $V_{\alpha\beta}$ defines the perturbation matrix.       

The resonant state expansion represents a generalization of the Brillouin-Wigner perturbation theory for non-Hermitian systems. Therefore, numerical accuracy is determined by the size of the basis set $N$. We choose the basis in such a way that for a given orbital number $l$, azimuthal number $n$ and parity we select all resonant states with frequencies lying inside the circle $|\omega R / c|<M$, where $R$ is the radius of the sphere that describes the cylinder.  Since the perturbation $V_{\alpha\beta}$ conserves the axial symmetry and up-down vertical symmetry, we study problem for each azimuthal index $n$ and each parity $p$ independently. 

\section{Estimation of efficiency of second harmonic generation}

In this section we derive rough estimation of second harmonic generation efficiency $\rho_{\rm SH}$. We work within the non-depleted pump approximation when the electric field of the second harmonic does not affect the modes at the fundamental frequency. We apply the temporal coupled mode theory to estimate the magnitude of linear and nonlinear response. 

Within this formalism we expand the incident field $\mathbf{E}_{\rm inc}e^{-i\omega t}$ over the contributions of independent channels $\mathbf{E}_p$ as 
\begin{equation}
\mathbf{E}_{\rm inc}(\mathbf{r})=\sum_p s_p^{(+)} \mathbf{E}_p(\mathbf{r}).
\end{equation}
Here each channel represents a mode of radiation continuum.

We assume that incident field frequency $\omega$ is  in the vicinity of frequency $\omega_{\rm FF}-i\gamma_{\rm FF}$ of one of the resonant states. Each channel couples to the resonant state with coupling strength $D^{\rm FF}_p$. Then the amplitude of the resonant state $a$ is evolving as
\begin{equation}
\frac{\partial a}{\partial t}=-i(\omega_{\rm FF}-i\gamma_{\rm FF})a + \sum_p D^{\rm FF}_p s_p^{(+)} .
\end{equation}
Next, we assume that we work in the resonant regime at the second harmonic frequency, which means that $2\omega$ is in the vicinity frequenecy $\omega_{\rm SH}-i\gamma_{\rm SH}$ of another resonant state.  Time dynamics of amplitude $b$ of this resonant state is governed by the nonlinear current source $J$, which is determined by the tensor of second order susceptibility $\chi^{(2)}_{ijk}$. Therefore,
\begin{equation}
\frac{\partial b}{\partial t}=-i(\omega_{\rm SH}-i\gamma_{\rm SH})b + a^2 J.
\end{equation}
The field $\mathbf{E}_{\rm SH}e^{-2i\omega t}$ radiated at second harmonic frequency can be also decomposed into contributions of independent channels  
\begin{equation}
\mathbf{E}_{\rm SH}(\mathbf{r})=\sum_p s^{(-)}_p \mathbf{E}_p(\mathbf{r}).
\end{equation}
Amplitudes $s^{(-)}_p$ can be determined via the coupled mode theory as
\begin{equation}
s^{(-)}_p= D^{\rm SH}_p b,
\end{equation}
where $D^{\rm SH}$ is the coupling strength between the second harmonic resonant state and the radiation continuum channels. 

Radiated power at SH can be calculated as
\begin{equation}
P_{\rm SH}=\frac{c}{8\pi k^2}\sum_p |s^{(-)}_p|^2
\end{equation}

Finally, the intrinsic second harmonic generation efficiency depends on $P_{\rm SH}$  as (see Eq.1 of the main text)
\begin{equation}
\rho_{\rm SH} = \frac{P_{\rm SH}}{\eta^2 P_{\rm FF}^2} =\frac{c}{8\pi k^2} \frac{1}{\eta^2 P_{\rm FF}^2} \sum_p |D^{\rm SH}_p|^2 |b|^2.
\label{eq:1}
\end{equation}
Substituting expressions for amplitudes $a$ and $b$ into Eq.~\ref{eq:1} we get
\begin{multline}
\rho_{\rm SH} = \frac{c}{8\pi k^2}\frac{1}{\eta^2 P_{\rm FF}^2} \left[\sum_p |D^{\rm SH}_p|^2\right] \times \\
\frac{J^2}{(2\omega-\omega_{\rm SH})^2+
\gamma_{\rm SH}^2}\left[\frac{|\sum_{p^{'}} D^{\rm FF}_{p^{'}} s_{p^{'}}^{(+)}|^2}{(\omega-\omega_{\rm FF})^2+\gamma_{\rm FF}^2}\right]^2.
\label{eq:2}
\end{multline}

Next, we assume that only radiation channel $p_{1}$ at the fundamental frequency and one channel $p_2$ at the second harmonic frequency dominates. Using the energy consumption theorem and the reciprocity theorem one can show that
\begin{align}
&2\gamma_{\rm FF} = c|D^{\rm FF}_{p_1}|^2,\\
&2\gamma_{\rm SH} = c|D^{\rm SH}_{p_2}|^2.
\label{eq:10}
\end{align}

Therefore, Eq.~\ref{eq:2} is reduced to
\begin{multline}
\rho_{\rm SH} = \frac{1}{2\pi c^2k^2}\left[ \frac{J \left|s_{p_1}^{(+)}\right|^2}{\eta P_{\rm FF}}\right]^2 \times \\ \frac{\gamma_{\rm SH}}{(2\omega-\omega_{\rm SH})^2+\gamma_{\rm SH}^2}\left[\frac{\gamma_{\rm FF}}{(\omega-\omega_{\rm FF})^2+\gamma_{\rm FF}^2}\right]^2.
\label{eq:3}
\end{multline}

The coupling amplitude $\eta$ defined in the text can rewritten in terms of coefficients of the temporal coupling mode theory, but we omit the derivation since it is cumbersome. In result, 
\begin{equation}
\eta = \mathcal{N}_0A\frac{c\left|s_{p_1}^{(+)}\right|^2}{P_{\rm FF}},
\end{equation}
where $A$ is the geometrical cross section of the resonator and $\mathcal{N}_0$ is the numerical constant.

Finally, we arrive at Eq.4 of the main text
\begin{equation}
\rho_{\rm SH} = C\frac{\gamma_{\rm SH}}{(2\omega-\omega_{\rm SH})^2+\gamma_{\rm SH}^2}\left[\frac{\gamma_{\rm FF}}{(\omega-\omega_{\rm FF})^2+\gamma_{\rm FF}^2}\right]^2.
\label{eq:13}
\end{equation}

Here we introduce the normalization constant $C$ as
\begin{equation}
C =  \frac{J^2}{\mathcal{N}_1 c^4k^2A^2},
\label{eq:14}
\end{equation}
where $\mathcal{N}_1$ is the numerical constant.

For results of Fig.3 of the manuscript we used exact values of resonant state frequencies and normalization $C$ was found phenomenologically by fitting to the shape of red curve in Fig.3.

\end{document}